\documentstyle[multicol,epsbox,aps,graphicx]{revtex}

\begin{document}
\draft
\title{Chiral Kosterlitz-Thouless transition in the frustrated
Heisenberg antiferromagnet \\
on a pyrochlore slab
}
\author{Hikaru Kawamura and Takuya Arimori}
\address{Department of Earth and Space Science, Faculty of Science,
Osaka University, Toyonaka 560-0043,
Japan}
\date{\today}
\maketitle
\begin{abstract}
Ordering of the geometrically frustrated two-dimensional
Heisenberg antiferromagnet on a pyrochlore slab
is studied by
Monte Carlo simulations.
In contrast to the kagom\'e Heisenberg antiferromagnet,
the model
exhibits locally non-coplanar  spin
structures at low temperatures,
bearing nontrivial chiral degrees of freedom.
Under certain conditions, the model exhibits a novel
Kosterlitz-Thouless-type transition at a finite temperature
associated with these chiral
degrees of freedom.
\end{abstract}
\begin{multicols}{2}
\narrowtext

Magnetic ordering of geometrically frustrated antiferromagnets (AFs)
has attracted interest of researchers in 
magnetism\cite{Diep,Collins,Kawareview}.
In geometrically frustrated
AFs,  spins usually
sit on lattices made up of triangles or tetrahedra
as elementary units, and interact antiferromagnetically
with their neighboring spins. Intrinsic inability to simultaneously satisfy all
antiferromagnetic nearest-neighbor interactions on a triangle or on a
tetrahedron necessarily leads to macroscopic frustration. This makes the
spin ordering on these lattices a highly nontrivial issue.

Recently, interest has been
focused on the properties of the lattices  consisting of
{\it corner\/}-sharing triangles or tetrahedra, {\it e.g.\/},
the 2D kagom\'e lattice and the 3D pyrochlore
lattice\cite{Ramirezreview,Schifferreview,Harrisreview}.
Due to the looser coupling among
the frustrating units,
these systems often
remain paramagnetic down to low temperatures. 
Experimentally, however,
many of the geometrically frustrated magnets
regarded as typical
kagom\'e or pyrochlore AFs exhibit a
phase transition at a low but finite temperature,
quite often a spin-glass (SG)-like freezing
transition\cite{Ramirezreview,Schifferreview,Harrisreview}.

One of the best studied geometrically frustrated AFs is the
$S=3/2$ Heisenberg kagom\'e AF,
SrCrGaO (SCGO)\cite{Ramirezreview,Schifferreview}.
Experimentally, this material exhibits a
SG-like transition at a finite temperature $T=T_f$ as in many other
geometrically frustrated AFs, although $T_f$
is considerably
lower than the Curie-Weiss temperature of this material due to the strong
geometrical
frustration\cite{Ramirez,Broholm}.
In spite of extensive experimental and theoretical efforts,
the true nature of this SG-like transition of SCGO
has remained elusive.
Although SCGO has been regarded for some time
as a typical model compound of the
2D kagom\'e AF,
Monte Carlo (MC) simulations
performed for the antiferromagnetic Heisenberg model
on the 2D kagom\'e lattice failed to reproduce the SG-like
transition as experimentally observed in SCGO\cite{Chalker,Reimers}, 
suggesting that the modeling
of SCGO as the kagom\'e AF might be inadequate in capturing
some essential aspects of
this material. In fact,
the underlying
lattice structure of SCGO is not of a pure (single-layer)
kagom\'e lattice, but rather, of a
kagom\'e sandwich, or a
``pyrochlore slab''\cite{Ramirezreview,Schifferreview,Ramirez,Broholm}.
The structure of the
lattice is illustrated in Fig.1:
It consists of two
2D kagom\'e layers
which sandwiches the sparse triangular
layer in between.
Note that this lattice is obtained by slicing the 3D pyrochlore
lattice along the (111) direction into the slab geometry.
Inelastic neutron-scattering measurements have indicated that
the neighboring slabs are
magnetically well separated along the $c$-axis\cite{Lee}.
Hence, in modeling SCGO,
one may safely neglect the inter-slab interaction
and consider the 2D Heisenberg model on a pyrochlore-slab lattice.

\begin{figure}
\begin{center}
\includegraphics[scale=0.65]{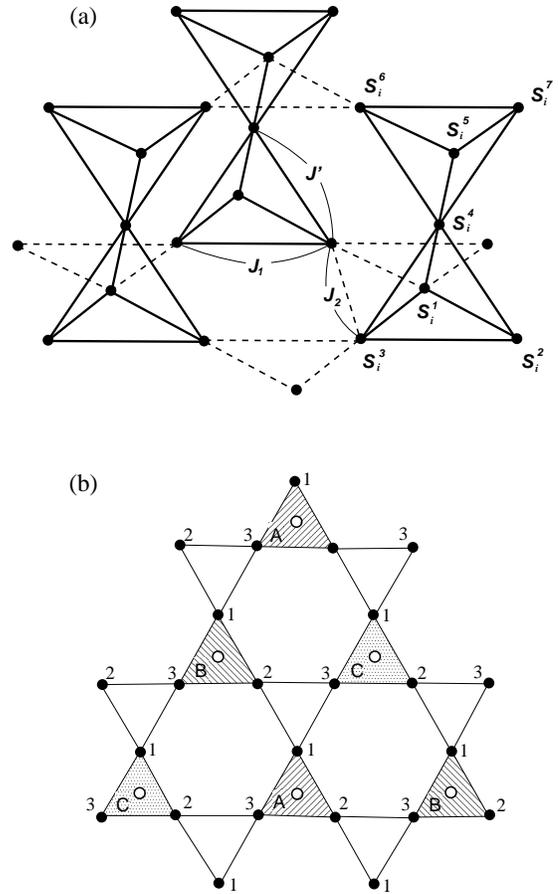}
\end{center}
\caption{
A pyrochlore-slab lattice. (a) The unit cell of the lattice consists of
two corner-sharing tetrahedra  illustrated  by the solid lines:
(b) The lower kagom\'e
layer consisting of the sites 1, 2 and 3, denoted by the
solid circle. The open circle denotes an apical site of
tetrahedron (site 4).
The upward triangles, shaded in the figure, correspond to
the bottom planes of tetrahedra, and are grouped
into three types A, B and C, each forming the three
inter-penetrating triangular sublattices.
}
\end{figure}

The  purpose of the present Letter is to study the ordering properties
of the antiferromagnetic classical Heisenberg model on a pyrochlore slab
by means of
MC simulations, and to examine whether some new
features which are different from  those
of the well-studied pure kagom\'e Heisenberg AF would arise,
possibly due to the tetrahedron-based structure
of this lattice. In particular, we pay attention to the possible ``chiral''
properties of the model. ``Chirality'' is  a multi-spin quantity
representing the sense or handedness of the local non-coplanar spin structures
induced by spin frustration. It is defined for
three neighboring Heisenberg spins as a pseudo-scalar, $\chi ={\vec S_1}\cdot
{\vec S_2}\times {\vec S_3}$, so as to give a nonzero value if the three
spins make non-coplanar configurations but vanish  otherwise.
In the case of the pure kagom\'e Heisenberg AF, it has been known that
the spin structure selected at low temperatures
is a coplanar one with the vanishing
chirality\cite{Chalker,Reimers}.
In sharp contrast to this, we  show in the case of the
pyrochlore-slab Heisenberg AF that
the spin structure stabilized at low temperatures is a
{\it non-coplanar\/}
one sustaining the nontrivial chirality.
These nontrivial chiral degrees of freedom
exhibit a novel thermodynamic
phase transition at a finite temperature without accompanying the
order of Heisenberg spins.

The unit cell of the pyrochlore-slab lattice may
be taken as two corner-sharing tetrahedra containing seven
sites, as illustrated  by the solid lines in Fig.1(a).
These unit cells containing seven sites, each
numbered from 1 to 7, are arranged forming the 2D
triangular lattice of spacing $2d$, $d$ being the lattice constant of the
kagom\'e layers.
In Fig.1(b), we show the lower kagom\'e layer
consisting of the sites 1, 2 and 3.
Note that the upward triangle
in Fig.1(b)
corresponds to the bottom plane of tetrahedron,
while the downward triangle is not a part of any tetrahedron.
These upward triangles are further grouped into three types,
$A, B$ and $C$, each forming triangular sublattices of spacing
$2\sqrt 3d$.

We consider the classical Heisenberg Hamiltonian,
\begin{equation}
{\cal H} = \sum_{\langle ij\rangle}J_{ij}
\vec S_i \cdot \vec S_j
+J^{'}\sum_{\langle mn\rangle }^{}
\vec S_m \cdot \vec S_n,
\end{equation}
where the antiferromagnetic interaction $J_{ij}$ on the kagom\'e layers
is assumed to
work between the nearest-neighbor (nn) and the
next-nearest-neighbor (nnn) pairs, with $J_{ij}=J_1$ and $J_2$,
respectively, while
$J^{'}$ is the antiferromagnetic nn interaction
between the kagom\'e layers and the triangular layer.
The variable $\vec S_i$
is a three-component unit vector
at the $i$-th site.
While we have made simulations
for several choices of $J_2/J_1$ and $J'/J_1$,
we report here on our results for the case of antiferromagnetic
nnn coupling $J_2>0$, because our model turns out to exhibit most
interesting transition behavior in this case.

Since the present model is  highly frustrated
showing very slow relaxation at low temperatures,
we combine the heat-bath MC method
with the temperature-exchange technique to facilitate efficient
thermalization\cite{HN}.
Simulations are made for a pyrochlore slab  with
$N=7\times N_s$ ($N_s=L\times L$) spins
with  $L=$6, 12, 18, 24 and 30.
Periodic boundary conditions are employed.

Now, we present our Monte Carlo results for the particular case of
$J_2/J_1=0.5$ and $J'=J_1$. In Figs.2(a) and (b),
we show the specific heat and
the uniform susceptibility, respectively.
The specific-heat data exhibit double peaks,
at $T=T_{p1}\simeq 0.29J_1$ and  at $T=T_{p2}\simeq 0.09J_1$.
Their size dependence
reveals that the peak heights eventually saturate
with $L$, suggesting that
both peaks are non-divergent ones.
Meanwhile, the  susceptibility data
exhibit a clear cusp-like anomaly
at $T/J_1\simeq 0.085$
close to  the lower specific-heat peak.

\begin{figure}
\begin{center}
\includegraphics[scale=0.7]{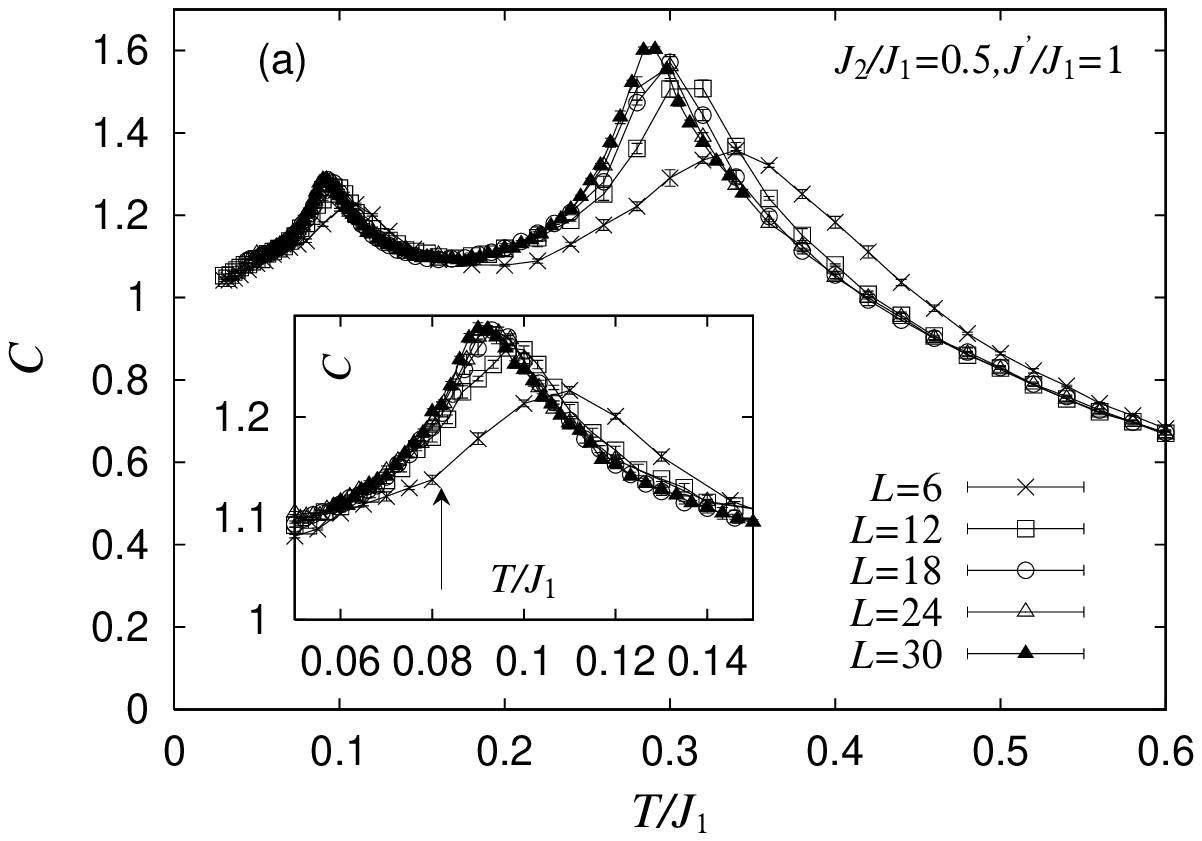}
\includegraphics[scale=0.7]{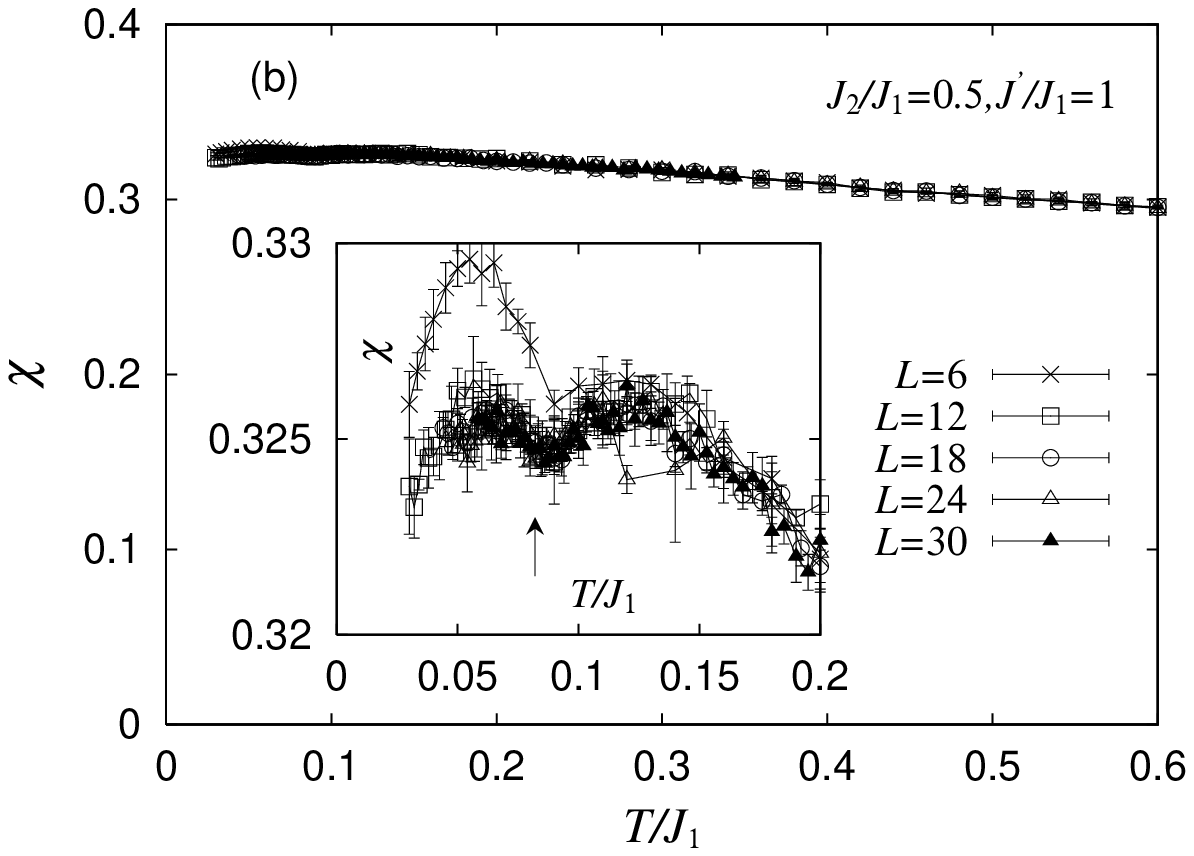}
\end{center}
\caption{
The temperature and size dependence of (a) the specific heat
and (b) the uniform susceptibility, per spin.
The insets are
magnified views of the low-temperature region. The arrow indicates
the transition point.
}
\end{figure}

One generally expects that, in a fully isotropic Heisneberg model in
2D like our model, there should be
no long-range order (LRO) of Heisenberg spins, nor a
finite-temperature transition occurring
in the spin sector.
We have confirmed this expectation
by calculating various order parameters
and Binder ratios associated with the spin order,
{\it e.g.\/}, the so-called $q=0$ and  $\sqrt 3\times \sqrt 3$ orders
on the kagom\'e layer\cite{Chalker,Reimers}.
We also find that
the higher specific-heat peak is correlated with the development
of the $q=0$ short-range order of Heisenberg spins on the kagom\'e
layers. In the present model with the antiferromagnetic $J_2$, 
the $q=0$ mode is selected over the $\sqrt 3\times \sqrt 3$ mode,
in contrast to the pure  kagom\'e Heisenberg AF where 
the $\sqrt 3\times \sqrt 3$ mode is selected\cite{Chalker,Reimers}.

\begin{figure}
\begin{center}
\includegraphics[scale=0.7]{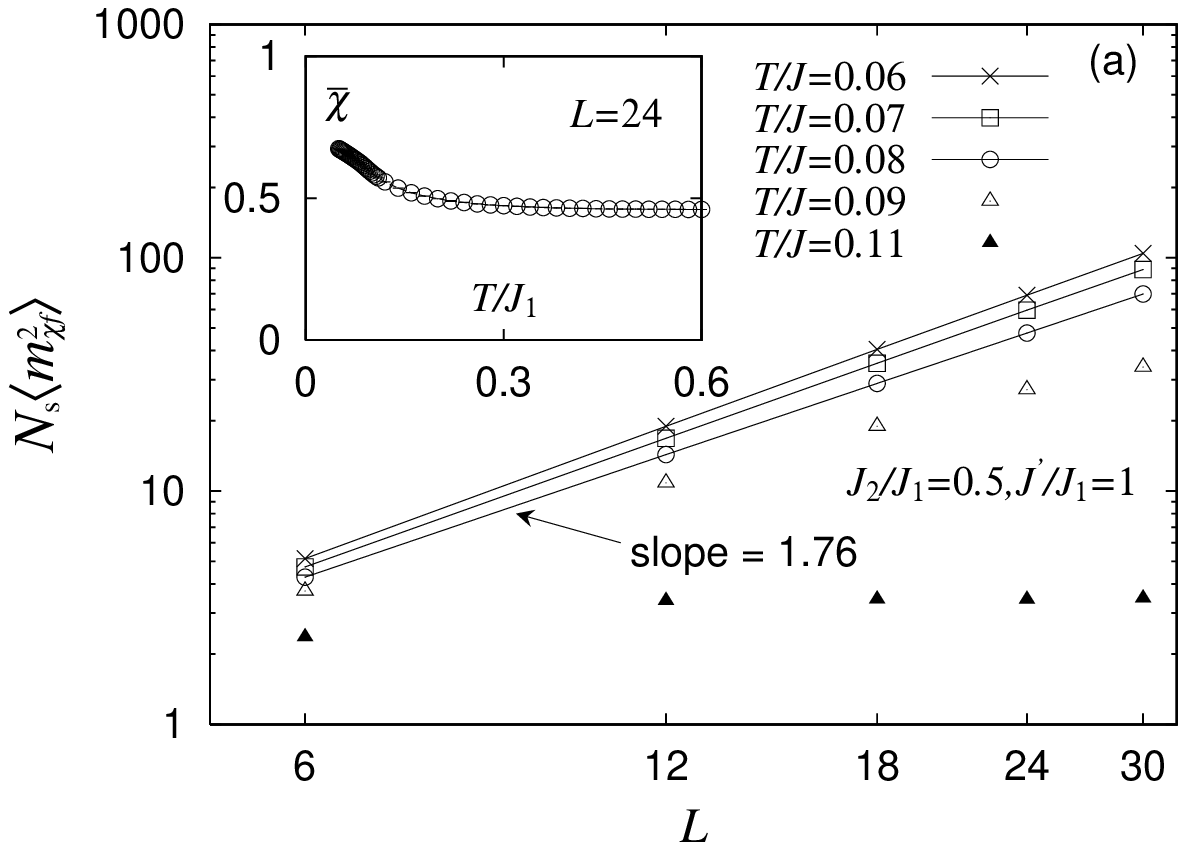}
\includegraphics[scale=0.7]{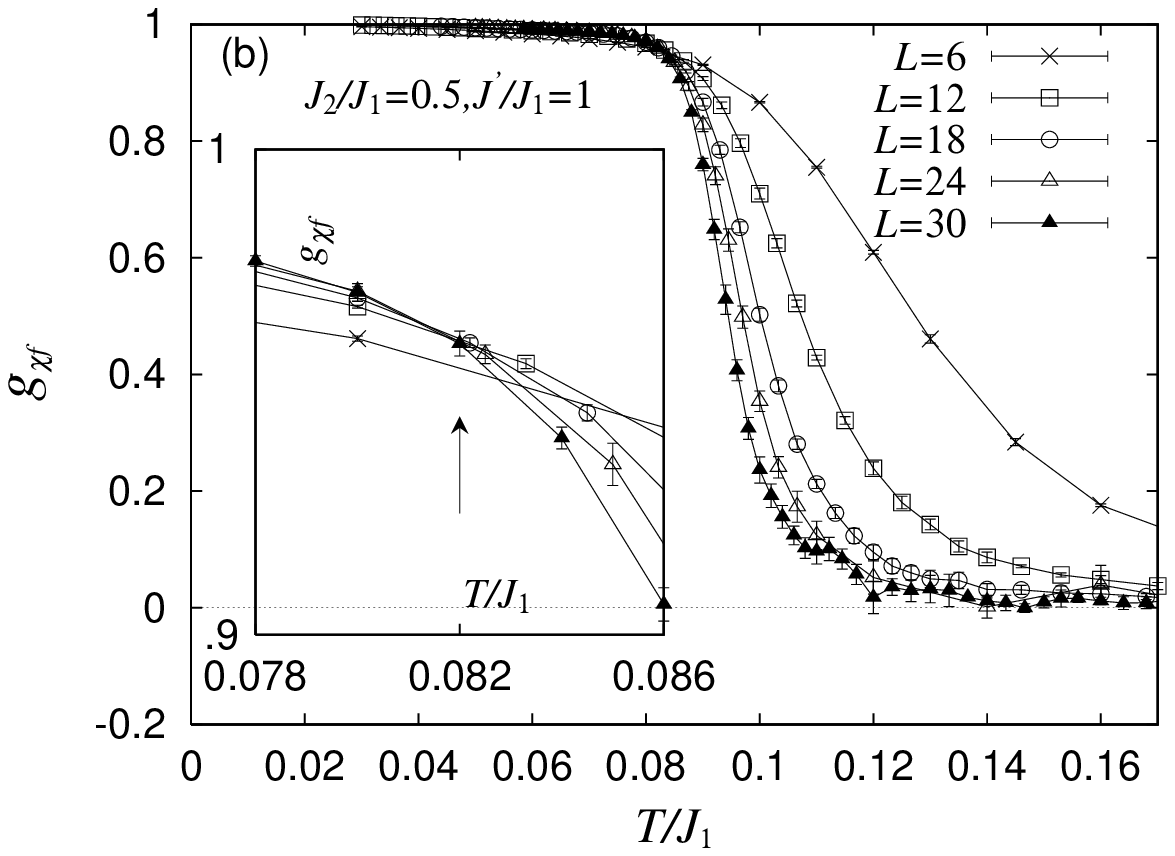}
\end{center}
\caption{
(a) Log-log plot of the $L$-dependence of the
chiral susceptibility associated with the $\sqrt 3\times \sqrt 3$
ordering at several temperatures. The inset
represents the temperature dependence of the mean local amplitude of the
chirality for $L=24$. The $L$-dependence of the data 
is negligible here, much smaller than the size of the symbols.
(b) The temperature dependence of the chiral
Binder ratio for various $L$. The inset is a magnified view of the transition
region.
The arrow indicates
the transition point.
}
\end{figure}

The chirality-related quantities are shown in Fig.3.
We define here the local chirality, $\chi_i$,
for the three spins 1, 2 and 3 located  at each upward
triangle on the lower kagom\'e layer.
In order to detect the local non-coplanarity of the spin structures,
we compute the mean
local amplitude of the chirality by $\bar \chi^2= (1/N_{\rm s})\sum _i
\langle \chi_i^2 \rangle$, 
where the summation is taken over all
upward triangles and
$\langle \cdots \rangle$ represents the thermal average.
The data shown in the inset of Fig.3(a) indicate
that in the $T\rightarrow 0$ limit $\bar \chi$
tends to a nonzero value,
meaning that  the spins belonging to tetrahedron
form locally non-coplaner structures
sustaining nontrivial chirality.
The appearance of the nontrivial chiral degrees of freedom in the
present model is
in sharp contrast to the behavior of the pure kagom\'e Heisenberg AF.

In order to detect the possible ordering of these chiralities,
we calculate the ferrimagnetic (or staggered)
chiral order parameter $m_{\chi f}$, defined by
\begin{eqnarray}
m_{\chi f}^2 &=& (m_{\chi}^A)^2+(m_{\chi}^B)^2+(m_{\chi}^C)^2
 \nonumber \\
&-&m_{\chi}^Am_{\chi}^B-m_{\chi}^Bm_{\chi}^C-m_{\chi}^Cm_{\chi}^A,
\end{eqnarray}
\begin{equation}
m_{\chi}^A=\frac{3}{N_s}\sum_{i\in A}\chi_i,\ \
m_{\chi}^B=\frac{3}{N_s}\sum_{i\in B}\chi_i,\ \
m_{\chi}^C=\frac{3}{N_s}\sum_{i\in C}\chi_i.
\end{equation}
This quantity gives a nonzero value if the chirality exhibits a ferrimagnetic
order with $\sqrt 3\times \sqrt 3$ periodicity on  three sublattices
$A$, $B$ and $C$ in Fig.1(b).
The associated chiral Binder ratio may be defined by
\begin{equation}
g_{\chi f}=2-\frac{\langle m_{\chi f}^4\rangle}
{\langle m_{\chi f}^2\rangle^2}.
\end{equation}

In Fig.3(b), we show the calculated
chiral Binder ratio.
For smaller $L$, $g_{\chi f}$ for various $L$
tend to cross at $T=T_{\rm c}\simeq 0.082J_1$,
but for larger $L$ they tend to merge at $T\leq T_c$,
signaling the
occurrence of a phase transition of the chirality.
The estimated transition temperature $T_{\rm c}/J_1=0.082(2)$ is in rough
agreement with the susceptibility-cusp temperautre estimated above,
and is slightly below the lower specific-heat peak
temperature $T_{p2}$. In fact, there is no apprecialbe anomaly
in the specific heat just at
$T=T_{\rm c}$: See the inset of Fig.2(a).
A merging behavior of the Binder
ratio, without
the discernible specific-heat anomaly just
at $T_c$ but with a nondivergent peak slightly above $T_c$,
suggests that the
observed chirality transition might be of the
Kosterlitz-Thouless (KT)-type. 

In Fig.3(a), we show on a log-log plot  the $L$-dependence of
the chiral susceptibility $N_s\langle m_{\chi f}^2\rangle $ associated with
the ferrimagnetic chiral order.
While the data at higher temperatures
exhibit the behaviors characteristic of the disordered phase,
bending down toward some finite values,
those at $T\leq T_c$
lie on straight lines, exhibiting the behavior expected for the
KT-like phase with algebraically-decaying  correlations.
The estimated slope of
the plots, which should be equal to $2-\eta$
with $\eta $ being the critical-point decay exponent,
turns out to be $\eta =0.24\pm 0.01$ at $T=T_c$ and gradually
decreases with decreasing temperature below $T_c$.
This again indicates that the observed chiral transition 
at $T=T_c$ is the KT-type transition.

Since the chirality here is an Ising like quantity sitting on
the triangualr lattice,
there is a close
similarity between the chirality ordering of the present model and
the ordering of the 2D Ising model on the triangular lattice.
Indeed, when the triangular Ising model possesses the antiferromagnetic nn
interaction and ferromagnetic nnn interation, the model
is known to exhibit a
$\sqrt 3\times \sqrt 3$
order via the KT-type transition characterized by
the exponent $\eta =1/4$\cite{Landau,Takayama,Fujiki,Miyashita}.
Since the chirality ordering of the present model with $J_2>0$
is essntially the staggered one,
it would be no surprise
that the chirality ordering here is of the KT-type. To the authors' knowledge,
our present finding is the first case of
{\it chiral\/} KT transition without accompanying the spin order.

It should be noticed that the
triangular Ising AF with the ferromagnetic
nnn interaction
exhibits another phase transition with further decreasing temperature,
into the low-temperature phase with a finite
LRO.\cite{Landau,Takayama,Fujiki,Miyashita}
The exponent $\eta $
at this second  transition is believed to be
$1/9$\cite{Landau,Miyashita,Jose}. In view of this,
we search for this second transition into the
long-range-ordered state in our present model, but with
negative result.
Presumably, severe frustration inherent to the present model might
hinder the
onset of the true LRO.

     Although we have so far considered the case where the inter-plane
interaction $J'$ is equal to the intra-plane nn interaction $J_1$,
such equality is not expected
in general in real experimental systems. In order to examine
the possible effect of varying the inter-plane coupling,
we also performed similar calculations for the $J'=0.5J_1$ case,
to get qualitatively similar results as the $J'=J_1$ case. We thus conclude
that the $J'$ value is irrevant to the chiral KT transition observed here.
Further detailed results of our MC simulations, including the results
for other choices of the nnn interaction
$J_2$, will be published elsewhere\cite{Arimori}. 
For the cases of the {\it vanishing\/} and {\it ferromagnetic\/} 
nnn interaction $J_2\leq 0$, we find that,
although the local spin structure is non-coplanar sustaining nontrivial 
chirality, the chiral KT transition does not take place there.

Finally, we wish to refer to the possible experimental implication of
our present results. We emphasize that
the present model is expected to capture
essential geometrical ingredients of SCGO, and indeed,
seems to accout for
some of the experimental features, {\it e.g.\/},
the occurrence of a finite-temperature
phase transition above all,
the cusp-like anomaly observed
in the susceptibility\cite{Ramirez},
or the existence of a broad specific-heat peak
slightly above $T_c$\cite{Ramirez}.
However, we mention that
some other experimental features
remain unexplained. For example, experimentally,
the transition is characterized by
the negative divergence of the nonlinear susceptibility
and the onset of magnetic irreversibility, {\it e.g.\/}, a
notable difference between
the field-cooled  and zero-field-cooled
magnetizations\cite{Ramirez}.
We could not reproduce these features in our simulations.

Although further work is clearly
required to fully account for
the experimental observation on SCGO,
our present finding of the
chiral KT transition in the pyrochlore-slab Heisenberg AF
might  hopefully give a new perspective in the studies of
SCGO, and geometrically frustrated AFs in general.

The numerical calculation was performed on the Hitachi SR8000 at the
supercomputer center, ISSP, University of Tokyo.

\end{multicols}


\begin{thebibliography}{99}



\bibitem{Diep} {\it Magnetic Systems with Competing Interaction\/}
ed. H.T. Diep, World Scientific (1994).


\bibitem{Collins} M. Collins and O.A. Petrenko, Can. J. Phys. {\bf 75}, 605
(1997).


\bibitem{Kawareview} H. Kawamura, J. Phys. Condes. Matter {\bf 10},
4707 (1998).


\bibitem{Ramirezreview}
A. P. Ramirez, Ann. Rev. Mater. Sci. {\bf 24}, 453 (1994).


\bibitem{Schifferreview} P. Schiffer and A. P. Ramirez,
Comments Cond. Mat. Phys. {\bf 18}, 21 (1996).


\bibitem{Harrisreview} M.J. Harris and M.P. Zinkin, Int. J. Mod.
Phys. B {\bf 10}, 417 (1996).


\bibitem{Ramirez}
A. P. Ramirez, G. P. Espinosa and A. S. Cooper,
Phys. Rev. Lett. {\bf 64}, 2070 (1990); Phys. Rev. B {\bf 45}, 2505 (1992).


\bibitem{Broholm}
C. Broholm, G. Aeppli, G. P. Espinosa and A. S. Cooper,
Phys. Rev. Lett. {\bf 65}, 3173 (1990).


\bibitem{Chalker}
J. T. Chalker, P. C. W. Holdsworth and E. F. Shender,
Phys. Rev. Lett. {\bf 68}, 855 (1992).



\bibitem{Reimers}
J. N. Reimers and A. J. Berlinsky,
Phys. Rev. B {\bf 48}, 9539 (1993).



\bibitem{Lee}
S. -H. Lee, C.Broholm, G. Aeppli, T.G. Perring, B. Hessen and A. Taylor,
Phys. Rev. Lett. {\bf 76}, 4424 (1996).



\bibitem{HN}
K. Hukushima and K. Nemoto, J. Phys. Soc. Jpn. {\bf 65}, 1604 (1996).



\bibitem{Landau}
D. P. Landau,
Phys. Rev. B {\bf 27}, 5604 (1983).


\bibitem{Takayama}
H. Takayama, K. Matsumoto, H. Kawahara and K. Wada,
J. Phys. Soc. Jpn. {\bf 52}, 2888 (1983).


\bibitem{Fujiki}
S. Fujiki, K. Shutoh, Y. Abe and S. Katsura,
J. Phys. Soc. Jpn. {\bf 55}, 3326 (1986).


\bibitem{Miyashita}
S. Miyashita, H. Kitatani and Y. Kanada,
J. Phys. Soc. Jpn. {\bf 60}, 1523 (1991).



\bibitem{Jose}
J. Jose, L. Kadanoff, S. Kirkpatrick and D. R. Nelson,
Phys. Rev. B {\bf 16}, 1217 (1977).


\bibitem{Arimori}
T. Arimori and H. Kawamura, in preparation.


\end{thebibliography}
\end{document}